\documentclass[twocolumn,floats,floatfix,showpacs,prl,superscriptaddress,nofootinbib]{revtex4-1}

\usepackage{graphicx,epsfig}
\usepackage{amssymb,amsmath,amsthm,amsfonts}
\usepackage{bm} 

\usepackage[linktocpage]{hyperref}
\usepackage[caption=false]{subfig}
\usepackage[usenames]{color}
\usepackage{url}

\def\dif{\textrm{d}}
\def\p{\partial}

\def\Lie{\mathcal{L}}
\def\G{\mathcal{G}}

\begin{document}

\title{Dynamical scalar hair formation around a Schwarzschild black hole}

\author{Robert Benkel}
\affiliation{School of Mathematical Sciences, University of Nottingham,
University Park, Nottingham, NG7 2RD, United Kingdom}

\author{Thomas P.~Sotiriou}
\affiliation{School of Mathematical Sciences, University of Nottingham,
University Park, Nottingham, NG7 2RD, United Kingdom}
\affiliation{School of Physics and Astronomy, University of Nottingham,
University Park, Nottingham, NG7 2RD, United Kingdom}

\author{Helvi Witek}
\affiliation{School of Mathematical Sciences, University of Nottingham,
University Park, Nottingham, NG7 2RD, United Kingdom}

\begin{abstract}
Scalar fields coupled to the Gauss-Bonnet invariant evade the known no-hair theorems and have nontrivial configurations around black holes. We focus on a scalar field that couples linearly to the Gauss-Bonnet invariant and hence exhibits shift symmetry. We study its dynamical evolution and the formation of scalar hair in a Schwarzschild background. We show that the evolution eventually settles to the known static hairy solutions in the appropriate limit. 
\end{abstract}

\maketitle

The recent direct observation of colliding black holes \cite{Abbott:2016blz,TheLIGOScientific:2016pea} marks the beginning of a new era in gravitational
physics. Gravitational wave observations promise to reveal new insights into the structure of black holes. According to general relativity (GR) their structure
should be remarkably simple and characterized by just three parameters:
their mass $M$,  angular momentum $J$, and  electromagnetic charge $Q$ \cite{Hawking:1971vc}. 
If observations were to disagree with this prediction the implications would be remarkable. For instance, this could reveal the existence of a new fundamental field that leaves an imprint on the structure of the black hole. This is precisely the scenario we consider here, and we focus on the case of scalar fields.

Apart from the obvious issue of having sufficiently accurate observations, there are also theoretical limitations to using black holes as probes for the existence of scalar fields. No-hair theorems dictate that minimally coupled, potentially self-interacting scalar fields have a trivial 
configuration around stationary, asymptotically flat black holes \cite{chase,Hawking1972,Bekenstein:1995un,Sotiriou:2012}. Since black holes are vacuum solutions, this result can be extended to large classes of nonminimally coupled scalar fields by virtue of conformal transformations and 
field redefinitions. Moreover, a similar result for the most general shift-symmetric scalar-tensor theory has been obtained in Ref.~\cite{Hui:2012qt} under the additional assumptions of staticity and spherical symmetry. This result extends trivially to slowly rotating solutions~\cite{Sotiriou:2013qea,Maselli:2015yva}.

One might be tempted to conclude that black holes cannot actually reveal the existence of scalar fields.
However, such a conclusion would not  only be premature but actually erroneous. 
No-hair theorems refer to stationary solutions, and
perturbations around such solutions can still carry a detectable imprint of a scalar field \cite{Barausse:2008xv}. 
Moreover, no-hair theorems rely on assumptions, and hence they can be circumvented.
For instance, hair can form due to the presence of matter around the black hole~\cite{Cardoso:2013fwa,Cardoso:2013opa};
if the scalar field does not respect the symmetries of the metric~\cite{Babichev:2013cya,Sotiriou:2013qea,Herdeiro:2014goa};
or if different asymptotics are considered~\cite{Jacobson:1999vr,Torii:2001pg,Horbatsch:2011ye};
see, e.g., Ref.~\cite{Sotiriou:2015pka} for a comprehensive discussion. 

Perhaps the most intriguing scenario, however, is one in which a theory  admits {\em only} hairy black hole solutions, even under the most stringent assumptions of staticity, spherical symmetry, asymptotic flatness, and complete absence of matter. This is the case when a scalar field is coupled
\ to the Gauss-Bonnet invariant 
$\G \equiv  R^{2} - 4 R_{ab} R^{ab} + R_{abcd} R^{abcd}$
either exponentially, i.e.,~$e^{\Phi} {\cal G}$ \cite{Kanti:1995vq} or linearly, i.e.,~$\Phi {\cal G}$ \cite{Sotiriou:2013qea}. 

If one further assumes that the scalar enjoys shift symmetry 
$\Phi \to \Phi + {\rm{constant}}$, then the following action, known as
Einstein-dilaton Gauss-Bonnet gravity,
\begin{equation}
\label{eq:Action2Couplings}
S \!=  \!\int \dif^{4} x \sqrt{-g} \left[ \frac{R}{16 \pi }
        + \mu \left(- \frac{1}{2} \nabla^{a} \Phi \nabla_{a}\Phi + \lambda  \Phi  \G \right) \right],
\end{equation}
has been shown \cite{Sotiriou:2013qea}  to evade the no-hair theorem of Ref.~\cite{Hui:2012qt}. Note that $\lambda$ and $\mu$ are coupling constants
and we employ units where $G=c=1$.
Variation of the action yields the field equations
\begin{align}
\label{eq:EoMsEdGBgeneralTen}
G_{ab} + \mu \lambda \kappa \G^{\rm{GB}}_{ab} = & 8\pi \mu T^{(\Phi)}_{ab} 
\,, \\
\label{eq:EoMsEdGBgeneralSca}
\Box \Phi = & - \lambda \G
\,,
\end{align}
where
\begin{align}
\label{eq:TmnSF}
T^{(\Phi)}_{ab} \equiv& \nabla_{a}\Phi \nabla_{b}\Phi 
        - \frac{1}{2} g_{ab} \nabla^{c}\Phi \nabla_{c}\Phi 
\,,\\
\G^{\rm{GB}}_{ab} = &
        g^{}_{g(a} g^{}_{b)j} \epsilon^{ghcd} \epsilon^{ijef} R_{cdef} \nabla_{h} \nabla_{i} \Phi\,.
\end{align}
In Eq.~\eqref{eq:EoMsEdGBgeneralSca} the scalar field is sourced by the Gauss-Bonnet invariant, which contains the Kretschmann scalar $R_{abcd} R^{abcd}$. 
This straightforwardly implies that  black holes are necessarily endowed with a nontrivial scalar configuration.

Static, spherically symmetric, asymptotically flat black hole solutions for this theory have been found in Ref.~\cite{Sotiriou:2014pfa}.
There is a unique, one-parameter family of solutions for which the scalar field is regular on the black hole horizon. Here we present the first exploration of the dynamical evolution of the scalar field and the formation of the nontrivial configuration that leads to black hole hair.

Modeling the collapse of a star and the formation of a black hole is not an easy task. Hence we resort to an approximation that provides significant simplifications and yet captures the key features of the problem. We neglect the effects the scalar field has on the geometry, and we study its dynamical evolution on a fixed spacetime. This can be formalized by taking the decoupling limit, $\mu \rightarrow 0$. In this limit, Eq.~\eqref{eq:EoMsEdGBgeneralTen} clearly reduces to Einstein's equations while Eq.~\eqref{eq:EoMsEdGBgeneralSca} remains entirely unaffected. Hence, the task at hand is to evolve Eq.~\eqref{eq:EoMsEdGBgeneralSca} on a spacetime background that is a GR solution. As a further approximation, we neglect the effect of matter and take this background to be the Schwarzschild solution. In a companion paper~\cite{longpaper} we drop this assumption and consider dynamical black hole formation described by the Oppenheimer-Snyder collapse~\cite{Oppenheimer:1939ue} in the background. However,  the simple case examined here turns out to be a remarkably good approximation for the scalar's dynamical behavior.

The first task is to determine what we expect to be the end point of our evolution. That is, we 
present the most general static, spherically symmetric, asymptotically flat solution to Eq.~\eqref{eq:EoMsEdGBgeneralSca}, which we will later compare to the late time behavior of our numerical simulations. 
We find it convenient to use isotropic coordinates, in which the Schwarzschild metric has the form
\begin{align}
\label{eq:SchwarzschildIsotropic}
\dif s^{2} = & - \alpha^{2}_{\rm{S}} \dif t_s^{2} + \psi^{4} \left( \dif \rho^{2} + \rho^{2} \dif\Omega^{2} \right)
\,,
\end{align}
where $\psi = 1 + M/(2\rho)$ is
the conformal factor  and $\alpha^{2}_{\rm{S}} =( M-2\rho )^{2} / ( M+2\rho )^{2}$ is the square of the lapse function.
The Schwarzschild areal radius coordinate $\bar{r}$ is related to $\rho$ by $\bar{r} = \psi^{2}\, \rho$,
so the horizon corresponds to $\rho=M/2$. 
Equation~\eqref{eq:EoMsEdGBgeneralSca} takes the form
\begin{align}
\label{eq:ScaEoMDecSchwarzschild}
\p_{\rho\rho}\Phi(\rho) 
- \frac{8\rho}{M^{2}-4\rho^{2} } \p_{\rho}\Phi(\rho)
+ \lambda \frac{48 M^{2}}{\rho^{6} \psi^{8}(\rho)}
= & 0
\,.
\end{align}
Generic solutions to this equation are not regular on the horizon. 
Imposing regularity there together with the asymptotic condition $\lim_{\rho\rightarrow\infty}\Phi = \Phi_{\infty}$ yields
\begin{align}
\label{eq:ScaSolSchwarzschild}
\Phi(\rho) = & \Phi_{\infty} 
        + \frac{2\lambda}{3M \rho^{3} \psi^{6} } \left( 4 M^{2} + 3 \rho M \psi^{2} + 3 \rho^{2}\psi^{4} \right)
\,.
\end{align}
Here $\Phi_{\infty}$ can be set to zero without loss of generality by exploiting shift symmetry. 
This solution is not new; after applying the coordinate transformation $\bar{r} = \psi^{2}\, \rho$ it matches the perturbative solution 
found in Refs.~\cite{Sotiriou:2013qea,Sotiriou:2014pfa,Yunes:2011we} at linear order in the coupling $\lambda$, or more rigorously, the dimensionless quantity $\lambda/M^2$. 
This is not surprising.
Working perturbatively in $\lambda$, at zeroth order Eq.~\eqref{eq:EoMsEdGBgeneralTen} is simply Einstein's equations 
coupled to a scalar field. Under the assumptions made here and in Ref.~\cite{Sotiriou:2013qea}, 
and in the absence of matter fields, no-hair theorems \cite{chase,Hawking1972,Bekenstein:1995un,Sotiriou:2012} apply; i.e., 
$\Phi$ needs to be in a trivial configuration, and the spacetime is described by the Schwarzschild solution. 
Then, to ${\cal O}(\lambda)$, $\Phi$ is determined by solving Eq.~\eqref{eq:EoMsEdGBgeneralSca} 
in this background; see Ref.~\cite{longpaper} for a more detailed discussion.

We now turn our attention to the evolution of the scalar field.
We commence by performing a spacetime decomposition of the background.
Although the metric~\eqref{eq:SchwarzschildIsotropic} is already written in $3+1$-form,
$t_s={\rm{constant}}$ slices do not actually penetrate the horizon, and hence $t_s$ is not a suitable time coordinate for our purposes.
We introduce a different slicing $\left(\Sigma_{t},\gamma_{ij}\right)$
with spatial metric $\gamma_{ab} = g_{ab} + n_{a} n_{b}$,
where $n^{a}$ is the unit timelike vector orthogonal to 
the hypersurfaces labeled by a time parameter $t$.
Then, the line element takes the form
\begin{align}
\label{eq:LineElement3p1}
\dif s^{2} = & g_{ab} \dif x^{a} \dif x^{b}
 \\
        = & - \left(\alpha^{2} - \beta^{k} \beta_{k} \right) \dif t^{2}
            + 2 \gamma_{ij} \beta^{i} \dif t \dif x^{j}
            + \gamma_{ij} \dif x^{i} \dif x^{j}
\,, \nonumber
\end{align}
where $\alpha$ and $\beta^{i}$ are, respectively, the lapse function and shift vector.
The extrinsic curvature of the constant $t$ hypersurfaces is 
$K_{ab} =  - \gamma^{c}{}_{a} \gamma^{d}{}_{b} \nabla_{c} n_{d}
        = - \frac{1}{2} \Lie_{n} \gamma_{ij}$
where $\Lie_{n}$ is the Lie derivative along $n^{a}$. 

Fixing our slicing would mean to provide expressions for $\alpha$, $\beta^i$ and $\gamma_{ij}$ in terms of $\alpha_S$ and $\psi$.
However, in practice we prefer to obtain the Schwarzschild background in a suitable foliation 
by evolving it numerically using puncture coordinates, i.e., the 1+log-slicing and the $\Gamma$-driver 
shift condition~\cite{Alcubierre:2002kk,Baker:2005vv,Campanelli:2005dd}. 
Note that the radial coordinate $r$ used in the numerics differs from the isotropic coordinate $\rho$, and the horizon lies at $r=r_{H}=1.09 M$. However, for $r> 10M$, $r$ and $\rho$ agree to at least $1\%$ accuracy.

In order to track the scalar field dynamics, we rewrite Eq.~\eqref{eq:EoMsEdGBgeneralSca} as a time-evolution problem,
\begin{align}
\label{eq:EvolSFdecoupling}
(\p_{t} - \Lie_{\beta}) \Phi = & - \alpha \Pi
\,,\\
(\p_{t} - \Lie_{\beta}) \Pi = & - \alpha \left(D^{i} D_{i} \Phi - K \Pi \right)
        - D^{i} \alpha D_{i} \Phi
        - \alpha\, \lambda \G
\,,\nonumber
\end{align}
where $\Pi = -\Lie_{n} \Phi$,
$D_{i}$ is the covariant derivative with respect to $\gamma_{ij}$,
and the invariant $\G$ reduces to the Kretschmann scalar of the Schwarzschild solution
since we are working in the decoupling limit.

We consider two sets of initial conditions $(\Phi,\Pi)|_{t=0}$:
\begin{align}
\label{eq:SFID1}
{\rm{ID~1:}} & \quad
\Phi_{0} = 0
\,,\quad
\Pi_{0}  = 0 
\,;\\
\label{eq:SFID2}
{\rm{ID~2:}} & \quad
\Phi_{0} = 0
\,,\quad
\Pi_{0}  = A_{0} e^{ \frac{(r-r_{0})^{2}}{\sigma^{2}} } \Sigma(\theta,\phi)
\,. 
\end{align}
In the first case the scalar starts out in a trivial configuration and with a vanishing time derivative. 
Its development is due to the sourcing by the Gauss-Bonnet invariant, so this case will demonstrate the inevitability of hair formation.
For the second case we choose the time derivative of the scalar to be a Gaussian shell with amplitude $A_{0}$, and width $\sigma$, centered around $r_{0}$. 
The angular configuration $\Sigma(\theta,\phi)$ is, in general, determined by a superposition of spherical harmonics $Y_{lm}$
such that the scalar field is real. 
Specifically, we consider the cases $\Sigma(\theta,\phi)=\Sigma_{00}\equiv Y_{00}$ and $\Sigma(\theta,\phi)=\Sigma_{11}\equiv Y_{1-1} - Y_{11}$.
We have evolved a number of configurations with amplitude $A_{0}/M=1$, and different widths $\sigma/M$ and locations $r_{0}/M$
of the Gaussian shell.
We set $M=1$, which can always be considered as a coordinate rescaling. 

We have implemented the field equations~\eqref{eq:EoMsEdGBgeneralTen} and~\eqref{eq:EoMsEdGBgeneralSca}
as part of the {\textsc{Lean}} code~\cite{Sperhake:2006cy},
which was originally based on the {\textsc{Cactus}} Computational toolkit~\cite{Cactuscode:web}
and the {\textsc{Carpet}} mesh refinement package~\cite{Schnetter:2003rb,CarpetCode:web},
and has now been adapted to the {\textsc{Einstein~Toolkit}}~\cite{EinsteinToolkit:web,Loffler:2011ay,Zilhao:2013hia}.
It has been extended to evolve additional fields in Refs.~\cite{Witek:2012tr,Okawa:2014nda,Zilhao:2015tya}.
We apply the method of lines to perform the evolutions,
where spatial derivatives are typically approximated by fourth-order finite differences
and we use the fourth-order Runge-Kutta time integrator. 
To simulate the background spacetime we employ the BSSN (Baumgarte-Shapiro-Shibata-Nakamura) formulation of Einstein's equations~\cite{Shibata:1995we,Baumgarte:1998te}
together with moving puncture coordinates~\cite{Alcubierre:2002kk,Baker:2005vv,Campanelli:2005dd}.
Our grid typically contains seven refinement levels, with the outer boundary located at $120M$ or $240M$ and 
resolution $h/M = 1.0$ on the outermost mesh. 
In order to access the discretization error we have performed a convergence analysis
using the additional resolutions $h_{{\rm{c}}}/M = 1.25$ and $h_{\rm{f}}/M = 0.75$.
We estimate the error to be 
$\lesssim 2\%$ in the waveforms after an evolution time of about $t/M \sim 200$.

To analyze the behavior of $\Phi$ we compute its energy density.
This can be split into
a canonical part 
$\rho_{\rm{SF}} =  \lambda^{-2} T^{(\Phi)}_{ab} n^{a} n^{b}$
and a part coming from the coupling to the Gauss-Bonnet invariant $\G$,
$\rho_{\rm{GB}} =  \lambda^{-1}\G^{\rm{GB}}_{ab} n^{a} n^{b}$.
We have deliberately introduced an extra $\lambda^{-2}$ factor in the definitions above, as it makes the 
corresponding quantities oblivious to the choice of $\lambda$. To see this consider the transformation $\Phi \to \lambda \Phi$.
At the level of the action~\eqref{eq:Action2Couplings}, this transformation allows one to effectively set $\lambda$ 
to $1$ by simply redefining $\mu$. 
This does not affect the process of taking the decoupling limit. Hence $\lambda$ becomes a redundant coupling at decoupling.
The same can be seen using  the field equations. At decoupling $\mu\to 0$, $\lambda$ and $\Phi$ are entirely absent 
from Eq.~\eqref{eq:EoMsEdGBgeneralTen}. The transformation $\Phi \to \lambda \Phi$ makes $\lambda$ drop out from Eq.~\eqref{eq:EoMsEdGBgeneralSca} as well. 
Clearly, instead of generating solutions for different values of the coupling constant, one can select a specific $\lambda$ and then obtain the remaining solutions simply by rescaling $\Phi$.
Hence, from now on we will just set the dimensionless coupling
$\lambda/M^{2}=1$.

\begin{figure}[t!]
\includegraphics[width=0.45\textwidth]{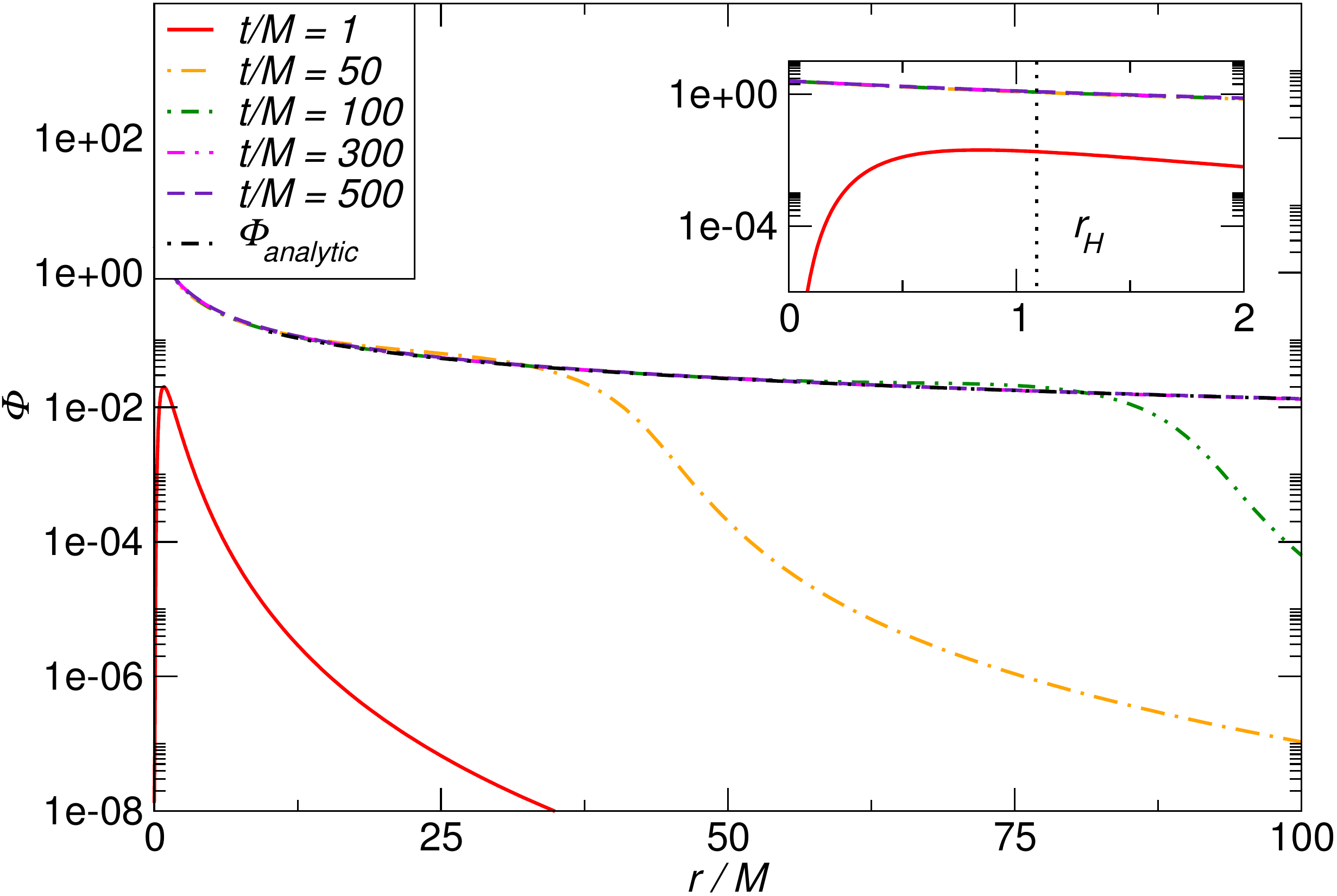}
\caption{\label{fig:IDSF0ScalarProfileRadial}
Radial profile 
at different instances of time for a scalar field that is initially vanishing and also has a vanishing time derivative.
We observe how the scalar field, sourced by the Kretschmann scalar, grows over time and approaches
the static, analytic solution in Eq.~\eqref{eq:ScaSolSchwarzschild} at late times.}
\end{figure}

\begin{figure}[h!]
\includegraphics[width=0.45\textwidth]{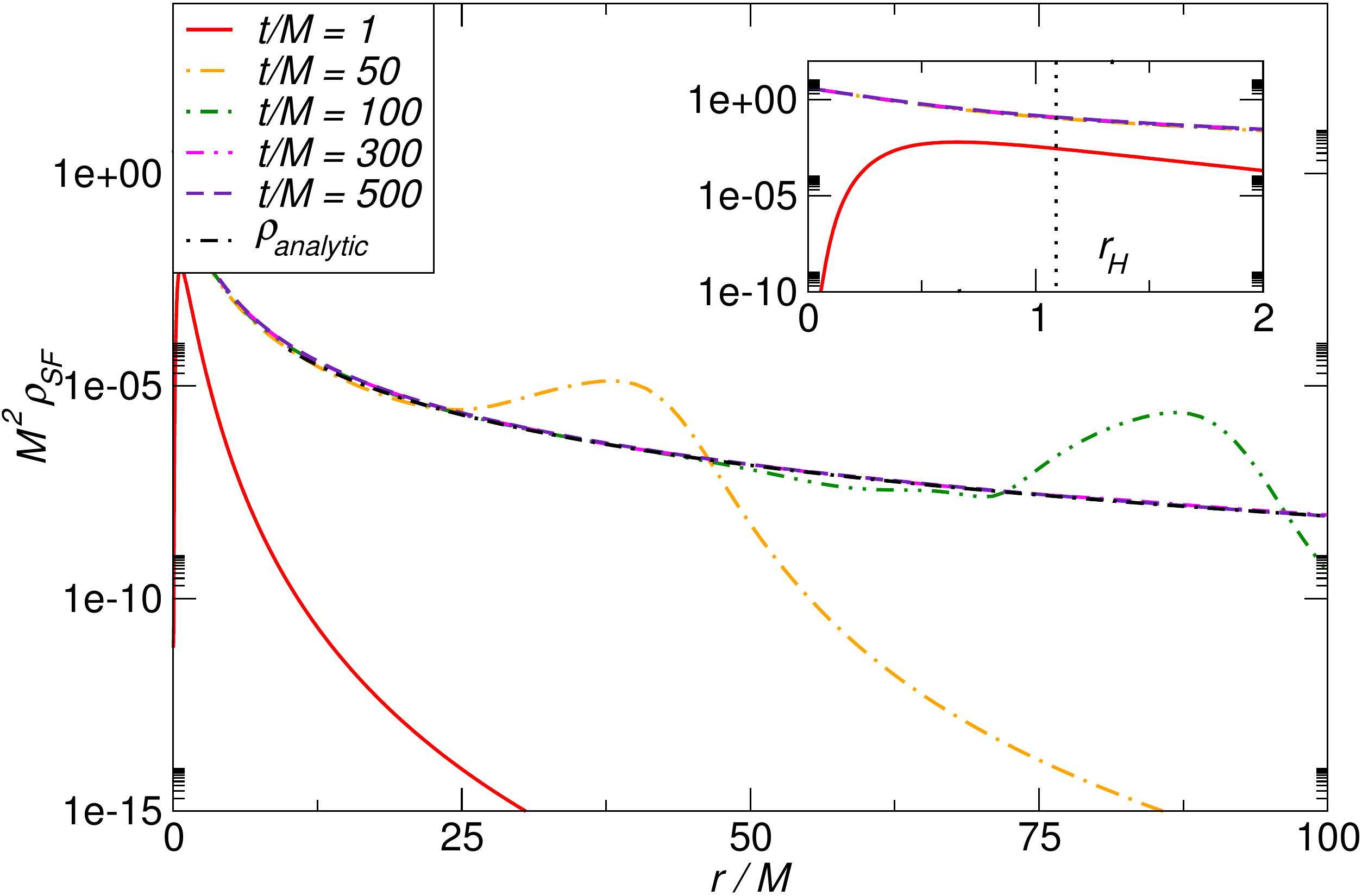}
\caption{\label{fig:IDSF0RhoProfileRadial}
Radial profile of the canonical energy density $\rho_{\rm{SF}}$ at different instances of time for a scalar field that is initially vanishing and also has a vanishing time derivative.
}
\end{figure}

\begin{figure}[htpb!]
\includegraphics[width=0.45\textwidth]{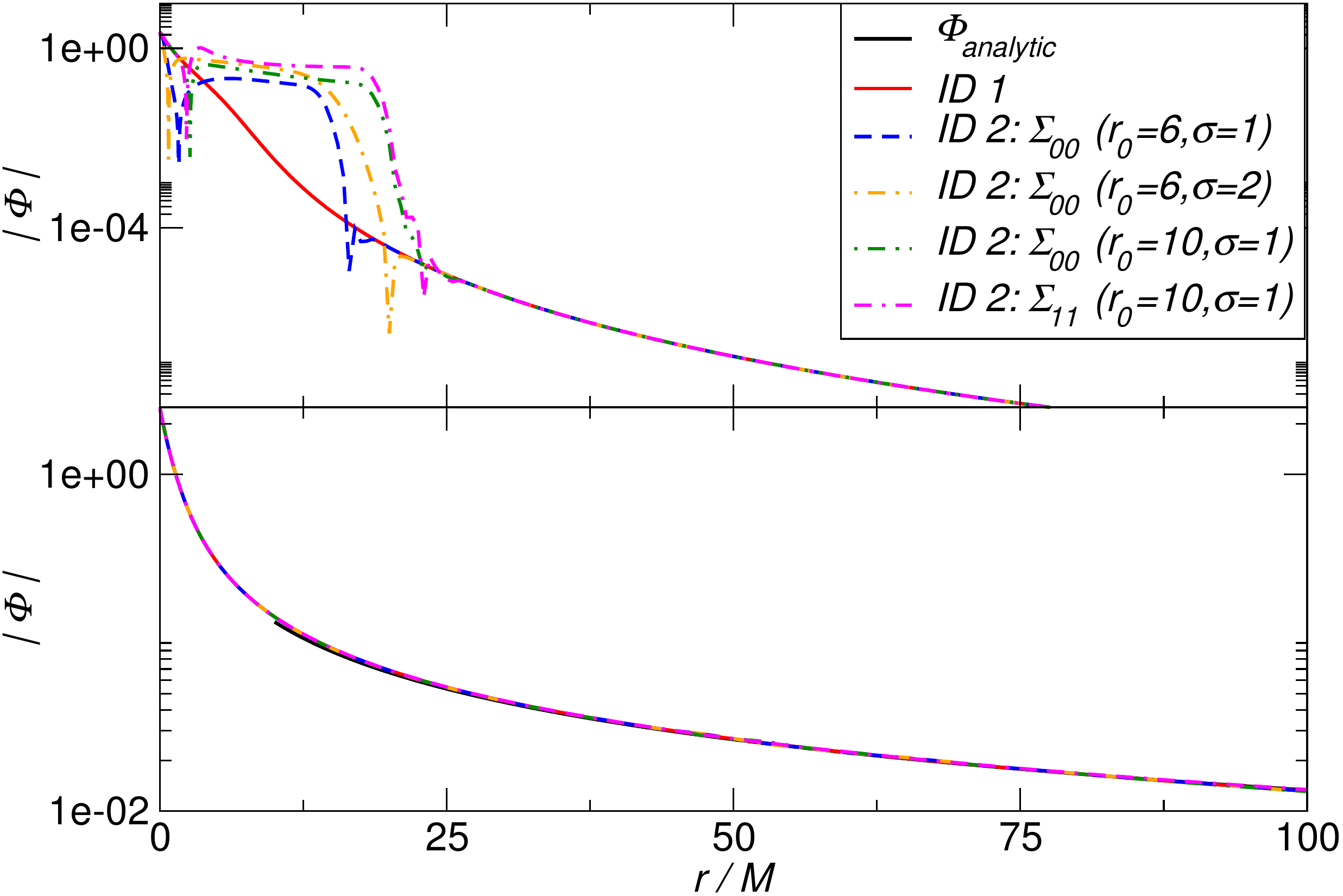}
\caption{\label{fig:Y004}
Radial scalar field profiles for different types of initial data at $t/M=10$ (top)
and at $t/M=300$ (bottom).
While the scalar field dynamics  are sensitive to the choice of initial data early in the evolution, in all cases the scalar settles to  the analytic solution in Eq.~\eqref{eq:ScaSolSchwarzschild} at late times.
}
\end{figure}

In Figs.~\ref{fig:IDSF0ScalarProfileRadial} and \ref{fig:IDSF0RhoProfileRadial} we show the radial profiles for $\Phi$ and $\rho_{\rm{SF}}$ at different instances during the evolution and for the initial conditions ID 1, given in Eq.~\eqref{eq:SFID1}. 
Note that $\rho_{\rm{SF}}$ and  $\rho_{\rm{GB}}$ scale differently with $\lambda$. In fact, $\rho_{\rm{SF}}/\rho_{\rm{GB}}\sim \lambda^{-1}$, and one can always choose $\lambda$ to be small enough so that  $\rho_{\rm{SF}}\gg \rho_{\rm{GB}}$ up to radii sufficiently close to the singularity. Hence, we prefer to show only $\rho_{\rm{SF}}$. 
It is clear that the scalar field and its energy density remain regular everywhere, and specifically in the vicinity of the horizon, throughout the evolution. Figure~\ref{fig:Y004}
depicts  the scalar's radial profile  for two instances of time and for various initial conditions of the type ID 2, given in Eq.~\eqref{eq:SFID2}.

In all cases the solutions relax to the known analytic, static, hairy scalar profile at late times.
Note that since $r$ and the isotropic  coordinate $\rho$ agree only at large radii, we have performed this comparison only for $r>10M$. However, assuming regularity on the horizon, the solution is unique \cite{Sotiriou:2014pfa}. Our numerical solutions are clearly regular at $r=r_H$, so agreement at large radii is sufficient.

\begin{figure}[t!]
\includegraphics[width=0.45\textwidth]{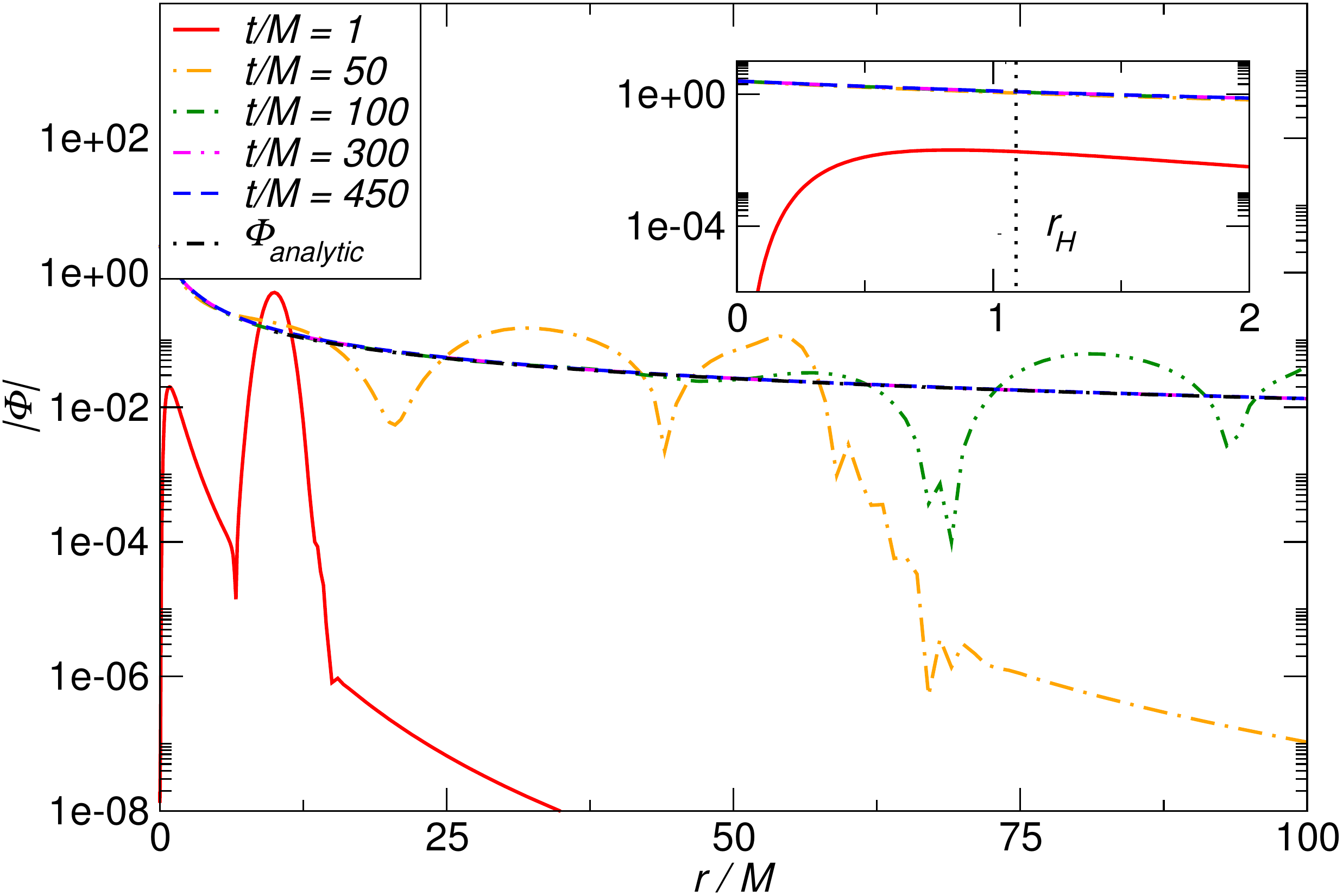}
\caption{\label{fig:IDSFY11ScalarProfileRadial}
Radial profile of an initially dipole scalar field shell with angular distribution $\Sigma_{11}$,
and parameters $A_{0}/M =1$, $r_{0}/M = 10$ and $\sigma/M = 1$ in Eq.~\eqref{eq:SFID2}.
Note, that we present the profile along the $\theta=0$ axis.
Despite the nonspherically symmetric initial configuration the field approaches the 
analytic solution in Eq.~\eqref{eq:ScaSolSchwarzschild} at late times,
after shedding its dipole moment as shown in Fig.~\ref{fig:Waveforms}.
}
\end{figure}

\begin{figure}[h!]
\includegraphics[width=0.45\textwidth]{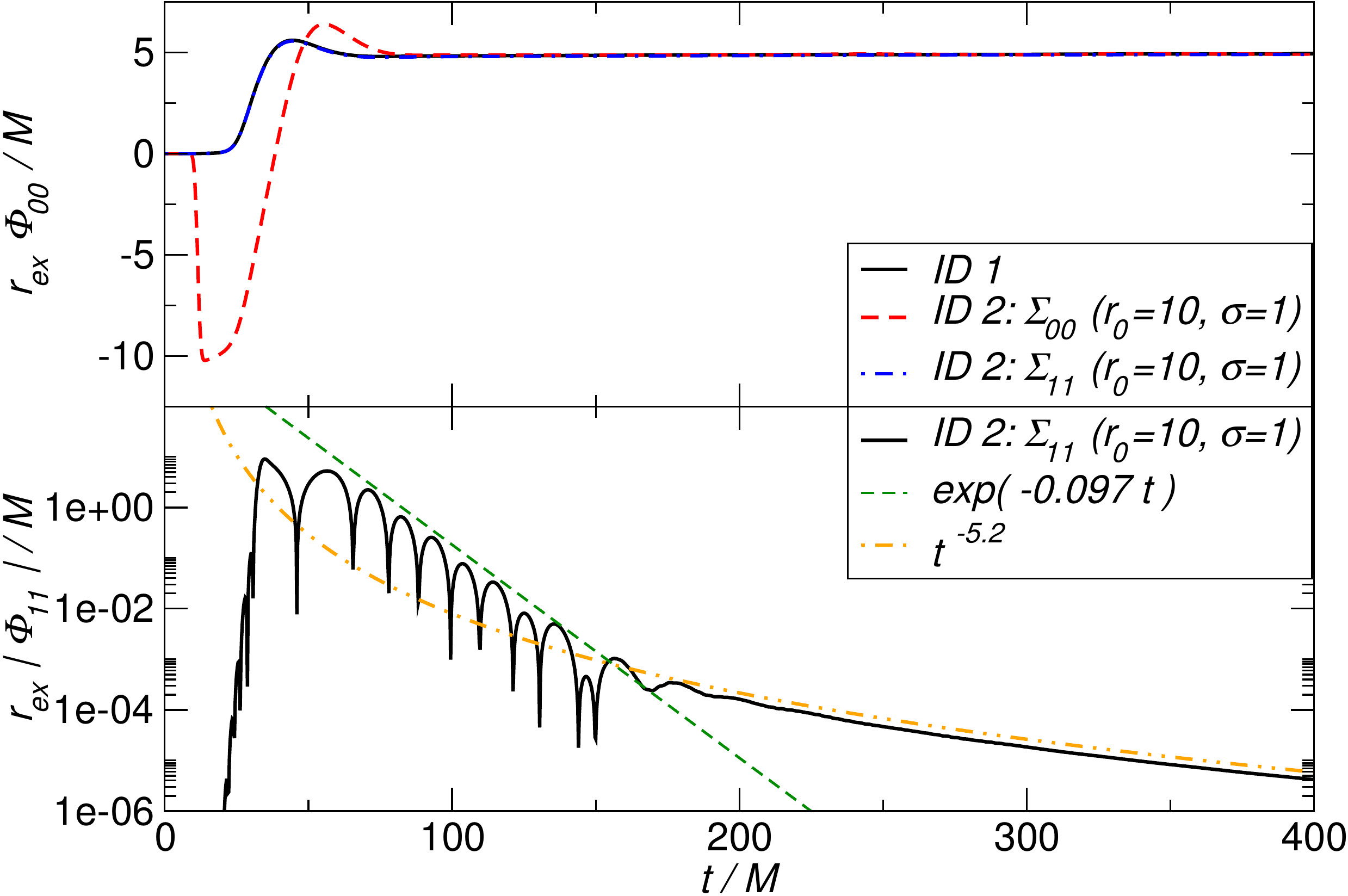}
\caption{\label{fig:Waveforms}
Scalar field waveforms, rescaled by the extraction radius $r_{\rm{ex}}/M = 20$.
Top panel:
Monopole mode  for the different types of initial data in Eqs.~\eqref{eq:SFID1} and~\eqref{eq:SFID2}
with $\Sigma_{00}$ or $\Sigma_{11}$.
Bottom panel:
Dipole mode for dipolar initial data, $\Sigma_{11}$ in Eq.~\eqref{eq:SFID2}.
In addition to the waveform (black solid line) itself we indicate the quasinormal mode damping $\sim e^{-0.097\,t}$
and late-time power-law tail $\sim t^{-5.2}$.
}
\end{figure}

We have mostly focused on initial data where $\Pi$ is initially a spherically symmetric cloud surrounding the central black hole, but we have also considered a dipolar initial profile $\Sigma_{11}$. The radial profile of the scalar for different time instances can be seen in Fig.~\ref{fig:IDSFY11ScalarProfileRadial}. In this case the scalar needs to shed away its dipole moment in order to relax to the known static configuration at later times. This is indeed the case, as shown in Fig.~\ref{fig:Waveforms}, where we present the $l=m=0$ (upper panel) and $l=m=1$ (lower panel) multipoles 
$\Phi_{lm}(t,r_{\rm{ex}}) = \int \dif\Omega \Phi(t,r_{\rm{ex}},\theta,\phi) Y^{\ast}_{lm}(\theta,\phi)$.
The monopole approaches the analytic solution at late times, 
while the dipole decays exponentially in time with complex frequency $M \omega = 0.2929 - \imath 0.097$
followed by the power-law tail at late times.
This is in excellent agreement, within $\lesssim 0.6\%$ and $\lesssim 4\%$ with predictions in GR~\cite{Berti:2009kk,Price:1971fb}.

In summary, we have performed numerical simulations for the evolution of a scalar field linearly coupled to the Gauss-Bonnet invariant in a Schwarzschild  background. The setup corresponds to the decoupling limit of theory \eqref{eq:Action2Couplings}. We focused on initial data for which the scalar vanishes initially and its derivatives either vanish or are given by a Gaussian shell.
In all cases, the scalar  eventually relaxes to the known static configuration of Eq.~\eqref{eq:ScaSolSchwarzschild}. 
This is the configuration of the known black hole solution obtained in Refs.~\cite{Sotiriou:2013qea,Sotiriou:2014pfa} 
(and also in Ref.~\cite{Yunes:2011we} in a different setup)
by working perturbatively in the coupling $\lambda$.


It is worth emphasizing that  there exists a two-parameter family of static, spherically symmetric, asymptotically flat black hole solutions for theory~\eqref{eq:Action2Couplings}, but generically the scalar diverges on the black hole horizon \cite{Sotiriou:2013qea,Sotiriou:2014pfa}. Regularity on the horizon singles out a one-parameter subclass in which the scalar charge is fixed by the black hole mass.
No regularity conditions have been imposed in our analysis, yet the scalar profile is  regular on the horizon and the scalar charge does acquire the desired value at late times. Hence, our simulations clearly demonstrate that this fix does not 
constitute tuning but instead arises naturally as the outcome of evolution. 
The scalar efficiently radiates away any extra charge or other features.

Our results provide strong evidence that this known one-parameter family of static black hole solutions is the end point of gravitational collapse in Einstein-dilaton Gauss-Bonnet gravity. The  sourcing of the scalar  by the Gauss-Bonnet invariant leaves a clear imprint in the evolution and makes it less sensitive to the choice of initial data. Remarkably, one can generate the same profile at late times even when both the scalar and its time derivatives are taken to vanish initially. This agrees with the intuition one gets from the static problem, where a nonvanishing Gauss-Bonnet invariant immediately implies a nonvanishing scalar, thereby leading to hairy black hole solutions. 

One could obtain more conclusive results by relaxing some of our simplifying assumptions. Still within the decoupling approximation, one can consider a background where a black hole forms dynamically. Indeed, in Ref.~\cite{longpaper} we have considered an Oppenheimer-Snyder background, and the results remain qualitatively unchanged. 
The next step would be to consider a more realistic description of a collapsing star. This is a particularly interesting scenario. In Ref.~\cite{Yagi:2015oca} it has been shown that a scalar that satisfies Eq.~\eqref{eq:EoMsEdGBgeneralSca} should have a vanishing monopole in stationary, asymptotically flat spacetimes without horizons. The monopole is instead nonzero for black hole spacetimes, so one expects a characteristic change in the scalar configuration upon the formation of a horizon. This can  be studied within the decoupling limit approximation. 
However, for a full description one needs to go beyond decoupling and take into account the backreaction of the scalar onto the metric. 

\acknowledgements
The research leading to these results has received funding from the European
Research Council under the European Union's Seventh Framework Programme
(FP7/2007-2013)/ERC Grant No.~306425 ``Challenging General
Relativity.'' 
HW acknowledges financial support by the European Union's H2020 research and innovation program under the Marie Sklodowska-Curie grant agreement No. BHstabNL-655360.
Computations were performed on the \textsc{minerva} HPC Facility at the University of Nottingham through Grant No. HPCA-01926-EFR
and on the \textsc{COSMOS} HPC Facility at the University of Cambridge,
operated on behalf of the STFC DiRAC HPC Facility and
funded by the STFC Grant No. ST/H008586/1, No. ST/K00333X/1, and No. ST/J005673/1, and on the \textsc{MareNostrum} supercomputer operated by the Barcelona Supercomputing Center and funded under Grant No. FI-2016-3-0006 ``New frontiers in numerical general relativity''. 



\end{document}